\documentclass[aps,prb,twocolumn]{revtex4}
\usepackage{graphicx}
\usepackage{amsmath}
\usepackage{bm}
\usepackage{xcolor}
\usepackage{epstopdf}
\begin{document}
\title{Double Andreev reflections and double electron transmissions in a normal-superconductor-normal junction based on type-II Weyl semimetal}
\author{Xue-Si Li$^1$}
\author{Shu-Feng Zhang$^2$}\email{sps_zhangsf@ujn.edu.cn}
\author{Xue-Rui Sun$^1$}
\author{Wei-Jiang Gong$^1$}\email{gwj@mail.neu.edu.cn}

\affiliation{1. College of Sciences, Northeastern University, Shenyang 110819, China\\
2. School of Physics and Technology, University of Jinan, Jinan, Shandong 250022, China }

\date{\today}

\begin{abstract}
We study the quantum transport behavior of a normal-superconductor-normal junction based on type-II Weyl semimetal, which is arranged in the tilting direction of the Weyl semimetal. We find that both the crossed Andreev reflection and normal reflection are forbidden, while there will be double Andreev reflections and double electron transmissions for the incident electron from the semimetal side. Andreev reflections and transmissions occur both in the retro and specular directions simultaneously, symmetric about the normal of the interface but with different amplitudes, depending on the angle and energy of incident electrons. These transport processes make the junction here quite different from that based on the normal metal or graphene. In addition, the differential conductance is studied for experimental signatures. We find that the conductance is almost unaffected by the chemical potential and it is enhanced with increasing junction length.
\end{abstract}
 \keywords{Type-II Weyl semimetal; Andreev reflection; Superconductivity}
\pacs{73.63.Kv, 71.70.Ej, 72.25.-b} \maketitle

\bigskip

\section{Introduction}
The interface between a normal and a superconducting material has been a widely used platform to realize exotic transport behaviors, especially in recent years when plenty of new materials with massless linear excitations or nontrivial topological properties are predicted and used to fabricate these interfaces \cite{AR-G1,AR-G2,AR-G3,AR-G4,AR-N,AR-IT1,AR-TS1,AR-WSM1}.
In a conventional normal metal-superconductor (NS) junction, there exists not only a normal reflection (NR) process but also an Andreev reflection (AR) process, in which an incident electron from the normal side is reflected back as a positive charged hole, and a Cooper pair forms in the superconductor (SC)\cite{also AR}. It is also known as retro AR, since the hole retraces almost the path of the incident electron. Following the successful fabrication of some new low-dimensional materials, novel AR phenomena have been reported from various aspects. One typical result is the specular AR, which has been discovered with the hole reflected along a specular path of the incident electron in the NS junction based on the graphene-like materials\cite{AR-G1,AR_WSM2}. And the perfect AR has been proposed and discovered in the NS junction of topological insulator\cite{perfect AR1,perfect AR2}. Besides, in the SNS junction of topological SCs, the fractional Josephson effect is allowed to come into being\cite{fractional josephson1,fractional josephson2}.

Very recently, the type-II Weyl semimetal (WSM) has been predicted and observed in several materials, e.g., $\mathrm{MoTe_{2}}$\cite{MoTe2_1,MoTe2_2}, $\mathrm{LaAlGe}$\cite{laAlGe}, $\mathrm{WTe_{2}}$\cite{WTe2_1,WTe2_2} and $\mathrm{Mo_{x}W_{1-x}Te_{2}}$\cite{MoWTe2_1,MoWTe2_2}. Like the type-I WSM, the low-energy excitations can be described by Weyl equation, but with a tilted anisotropic energy spectrum.
For the type-I WSM, the conduction and valence bands intersect at several Weyl nodes and thus the Fermi surface is point-like and the spectra around the nodes are coniclike~\cite{AR-G2}. However, the spectra of the type-II WSM are tilted by rotating around the Weyl nodes and there will exist electron and hole pockets near the line-like Fermi surface with a large density of states\cite{type-II1,type-II2,type-II3}. Therefore, the type-II WSMs will show a lot of interesting properties, in comparison with the type-I WSMs, such as the anomalous Hall effect, chiral anomaly and other intriguing properties\cite{anomalous Hall1,field-selective,anomalous Hall2,anomalous nernst,chiral anomaly,magnetic breakdown}.
Besides, the tilted energy bands of type-II WSM have opportunities to bring new transport mechanisms. According to the previous works, the phenomenon of double ARs has been predicted at the interface between the type-II WSM and its-based SC\cite{double}, where the retro and specular ARs coexist accompanied by the forbidden NR.
\par
In view of the AR result contributed by the type-II WSM, one can be sure that novel transport behavior will emerge in a normal-superconductor-normal (NSN) junction based on type-II WSM. Motivated by such a topic, in the present work we concentrate on the NSN junction arranged in the tilting direction of the type-II WSM and perform the discussion the AR properties, with the help of the scattering matrix method. As is known, in a conventional NSN junction of normal metal, four transport processes coexist, i.e., NR, normal electron transmission (ET), AR, and crossed AR, as shown in Fig.~\ref{fig:schematic} (a)\cite{NSN1,NSN2,NSN3,GSG1,GSG2,GSG3}, where the crossed AR process refer to the incident electron and the hole come from the different metal on both sides of the SC.
However, our study indicates that the NR and crossed AR are forbidden in the NSN junction based on the type-II WSM. Instead, the double ARs and double ETs happen simultaneously, including retro Andreev reflected ($A_1$), specular Andreev reflected ($A_2$), normal transmission ($T_1$), and the abnormal specular transmission ($T_2$), just as shown in Fig.~\ref{fig:schematic} (b). It shows that two ARs (ETs) have the same reflection (refraction) angle but with different amplitudes in general. In addition, the relationships between the four scattering processes and some controllable variables, e.g., the chemical potential, the incident angle and the junction length, are exhaustedly investigated in this paper. Also, the differential conductance is studied, and it has been found to have a larger magnitude considering the large momentum mismatch between the normal and SC region. Moreover, it is almost unaffected by the chemical potential and enhanced with the increase of junction length, which can be viewed as the obvious signatures for experimental observation.
\par
The rest of the paper is organized as follows. In Sec. II, we give the model Hamiltonian and introduce the phenomenon of double ARs and double ETs. In Sec. III, we calculate the amplitudes of these four transport processes with the scattering matrix method. In Sec. IV, we study the differential conductance. In Sec. V, we give a brief conclusion.

\section{The phenomenon of double Andreev reflections and double electron transmission}
\subsection{Model}
\begin{figure}
\begin{center}\scalebox{0.27}{\includegraphics{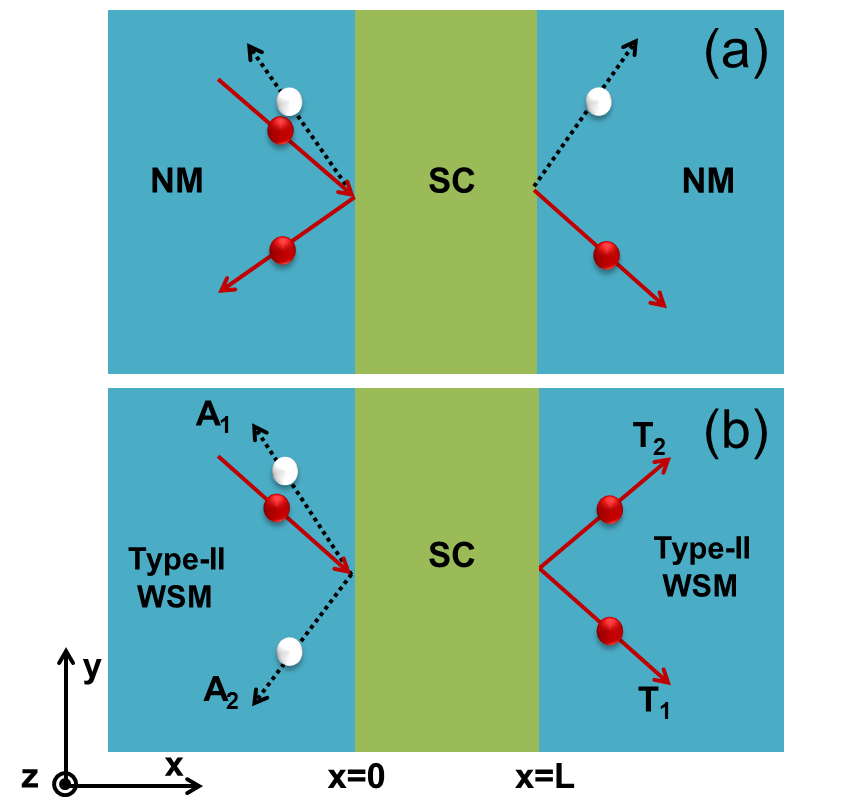}}
\caption{Schematic diagrams for reflection and transmission processes in the NSN junction based on (a) NM and (b) type-II WSM. There are four process in (a) : NR, normal ET, retro AR and crossed AR. However, there are no crossed AR and NR in (b). Instead, double ARs and double ETs occur. The red bullet denotes the electron, and the white bullet denotes the hole.
\label{fig:schematic}}
\end{center}
\end{figure}

The schematic of the NSN junction based on the type-II WSM is shown in Fig.~\ref{fig:schematic} (b). We model the type-II WSM with a most simple low-energy effective Hamiltonian which respects time reversal symmetry while breaks spatial inversion symmetry~\cite{hamiltonian,time reversal}. Near Weyl point $\bm{K_{0}}$, the Hamiltonian is written as\cite{double}
\begin{eqnarray}
H_{+}(\bm{k})=\hbar v_{1}k_{x}\sigma_{0}+\hbar v_{2}\bm{k}\cdot\bm{\sigma},
\end{eqnarray}
in which $\bf k$ is the wave vector measured from $\bm{K_0}$. $\bm{\sigma}=(\sigma_{x},\sigma_{y},\sigma_{z})$ denotes the Pauli matrix, $\sigma_{0}$ is the identity matrix. $v_2$ describes the Fermi velocity and $v_1$ determines the tilt along $x$ direction. It is a type-II WSM for $|v_1|>v_2$ while a type-I WSM for $|v_1|<v_2$.
The Hamiltonian near the $-\bm{K_{0}}$ node is related with that near $\bm{K_{0}}$ via time reversal symmetry, and it takes the form
\begin{eqnarray}
H_{-}(\bm{k})=-\hbar v_{1}k_{x}\sigma_{0}+\hbar v_{2}(k_{x}\sigma_{x}-k_{y}\sigma_{y}+k_{z}\sigma_{z}).
\end{eqnarray}
In the SC region, Cooper pairing occurs between the electrons near $\pm\bm{K_{0}}$. We consider the BCS pairing for simplicity. We use the electron and hole representations for quasiparticles near $\pm\bm{K_{0}}$ respectively. Then the Bogoliubov-de Gennes (BdG) Hamiltonian of the whole junction in real place can be written as\cite{BdG}
\begin{equation}\label{EqHBdG}
H_{BdG}=
\left[
  \begin{array}{ccc}
    H_{+}(\bm{k})-\mu(\bm{r}) & \Delta(\bm{r})\\
    \Delta^{*}(\bm{r}) & -H_{+}(\bm{k})+\mu(\bm{r})\\
  \end{array}
\right],
\end{equation}
where the wave vector has been replaced with ${\bf{k}} = -i\nabla_{\bf r}$, since translation invariance is broken along $x$-direction. $\mu(\bm{r})$ and $\Delta(\bm{r})$ are the chemical potential and superconducting order parameter, respectively, which can be expressed as
\begin{eqnarray}\label{eq:array}
&&\mu(\bm{r})=\left\{
\begin{array}{ll}
\mu&\text{if $x<0$, or $x>L$}\\
U&\text{if $0<x<L$}\\
\end{array}
\right.,\notag\\
&&\Delta(\bm{r})=\left\{\begin{array}{ll}
0&\text{if $x<0$, or $x>L$}\\
\Delta&\text{if $0<x<L$}\\
\end{array}\right..
\end{eqnarray}

\begin{figure*}
\begin{center}\scalebox{0.45}{\includegraphics{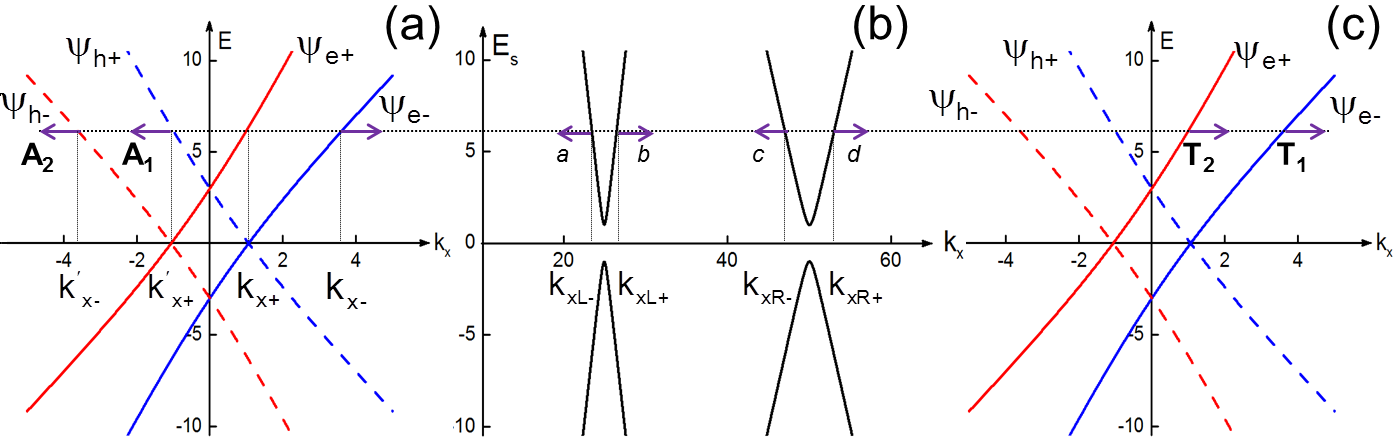}}
\caption{Energy spectra with finite $k_{y}$ and $k_{z}$ for (a,c) the type-II WSM and (b) SC region. In panel (a), the conduction/valence bands are colored with red/blue, and the solid/dashed lines denote electrons/holes. The arrows denote the direction of the incident, reflected and transmitted quasiparticles. The parameters : $k_{y}=3$, $k_{z}=0$, $v_{1}=3$, $v_{2}=1$, $\mu=0$, $\Delta=1$, $U=100$.
\label{fig_s2}}
\end{center}
\end{figure*}

The energy dispersion for quasiparticles in the SC region is written as
\begin{eqnarray}
   E_{S}(\emph{\textbf{k}})=\pm\sqrt{\Delta^{2}+(\hbar v_{1}k_{x}\pm\hbar v_{2}k-U)^{2}}.
\end{eqnarray}
The dispersion for eigenstates with $k_y=3,k_z=0$ is plotted in Fig.~\ref{fig_s2} (b) and it indicates that there are two right-moving and two left-moving modes for a fixed energy.
In the WSM region, superconducting order $\Delta$ vanishes and there are two electron modes and two hole modes with energy dispersions given as
\begin{eqnarray}
E_{e\pm}(\emph{\textbf{k}})&=& \hbar v_{1}k_{x}\pm  \hbar v_{2}k-\mu,\\
E_{h\pm}(\emph{\textbf{k}})&=&- \hbar v_{1}k_{x}\pm \hbar v_{2}k+\mu,
\end{eqnarray}
in which $k=\sqrt{k_x^2+k_y^2+k_z^2}$. $E_{e+}$ ($E_{h-}$) is the conduction band for electrons (holes), while $E_{e-}$ ($E_{h+}$) describes the valence band. The group velocity is the gradient of dispersion ${\bf{v}} = \frac{1}{\hbar}\nabla_{\bf{k}} E(\bf{k})$ and can be derived as
\begin{eqnarray}\label{EqVg}
  &&v_x^{e+} = v_1 + v_2 k_x/k, v_y^{e+} = v_2 k_y/k,\nonumber\\
  &&v_x^{e-} = v_1 - v_2 k_x/k, v_y^{e-} = -v_2 k_y/k,\nonumber\\
  &&v_x^{h+} = -v_1 + v_2 k_x/k, v_y^{h+} = v_2 k_y/k,\nonumber\\
  &&v_x^{h-} = -v_1 - v_2 k_x/k, v_y^{h-} = -v_2 k_y/k.
\end{eqnarray}
The $z$-component, which is missed here, shares a similar form as the $y$-component since the system is rotation invariant about $x$-axis. It is clear that both electron modes are right-moving while both hole modes are left-moving for type-II WSM with $|v_1|>v_2$, as is shown in Fig.~\ref{fig_s2} (a) and \ref{fig_s2} (c).

\subsection{Double ARs and double ETs}
Now we consider an incident electron from the left WSM region in the channel $\psi_{e-}$ with energy $E$ and wave vectors $k_y$ and $k_z$, shown in Fig.~\ref{fig_s2} (a). Because of the translation invariance, $k_y$ and $k_z$ are good quantum numbers and keep invariant in the transport process. Double ARs occur at the interface between the left WSM and SC regions. The incident electron will be retro reflected ($A_{1}$) or specular reflected ($A_{2}$) as a hole of modes $\psi_{h+}$ or $\psi_{h-}$, respectively. These two Andreev modes are symmetric about the normal of the interface in the incident plane with the same reflection angle $\theta_{h}=\arctan(\frac{v_2k_y}{\sqrt{(E-\mu)^{2}+(v_{1}^{2}-v_{2}^{2})k_{y}^{2}}})$ but different amplitudes depending on the incident angle and energy. Double ETs occur at the interface between the right WSM and SC regions. The incident electron transmits into the intraband and interband modes $\psi_{e-}$ and $\psi_{e+}$, respectively. The intraband/interband ET corresponds to the normal/specular ET with amplitude $T_{1}$/$T_{2}$.  The intraband mode is identical to the incident mode while the specular ET is symmetric with the normal ET, with the same angle $\theta_{e}=\arctan(\frac{v_2k_y}{\sqrt{(E+\mu)^{2}+(v_{1}^{2}-v_{2}^{2})k_{y}^{2}}})$.
On the other hand, we note that the NR and crossed AR are forbidden. These novel transport behaviors make this junction distinguished from the junction based on normal metal or graphene\cite{NSN1,NSN2,NSN3,GSG1,GSG2,GSG3}.

\section{amplitudes of double Andreev reflections and double electron transmissions}
In this section, we study in detail the amplitudes of each transport process of the double ARs and double ETs via the scattering matrix method. The effect of chemical potential and junction length is also investigated. In the numerical calculations, we set $v_1=2$, $v_{2}=1$, $\Delta=1$, $U=100$ and $\mu=0.5$ unless otherwise stated.
\subsection{Formalisms}
The amplitude of each transport process and the differential conductance can be solved by the scattering matrix method. Since this system is rotation invariant about $x$-axis, we set $k_z=0$ in the following discussion. Consider an electron incidents with energy $E$ and wave vector $k_y$. The wave-functions in the three regions can be expressed as
\begin{eqnarray}
&&\Psi_{\bf{I}}(\bm{r})= \Psi_{e-}(\bm{r})+r_{1}\Psi_{h+}(\bm{r})+r_{2}\Psi_{h-}(\bm{r}),\notag\\
&&\Psi_{\bf{II}}(\bm{r})= a\Psi_{a}(\bm{r})+b\Psi_{b}(\bm{r})+c\Psi_{c}(\bm{r})+d\Psi_{d}(\bm{r}),\notag\\
&&\Psi_{\bf{III}}(\bm{r})=t_{1}\Psi_{e-}(\bm{r})+t_{2}\Psi_{e+}(\bm{r}),
\end{eqnarray}
where I, II, III denotes the left, central, and right region, respectively. $\psi_{e\pm}$ ( $\psi_{h\pm}$ ) are the eigenvectors of two electron (hole)-modes and $\Psi_{a,b,c,d}$ are corresponding modes in the SC region. The expressions of these eigenvectors are given in the Appendix.
$r_{1/2}$ is the retro/specular AR coefficient, and $t_{1/2}$ is the normal/specular ET coefficients. $a$, $b$, $c$, and $d$ are coefficients of the qusiparticle modes in the SC region.
These coefficients are determined by the boundary conditions at the two interfaces,
\begin{eqnarray}
&&\Psi_{\mathbf{\uppercase\expandafter{\romannumeral1}}}(\bm{r})|_{x=0^{-}}= \Psi_{\mathbf{\uppercase\expandafter{\romannumeral2}}}(\bm{r})|_{x=0^{+}},\notag\\
&&\Psi_{\mathbf{\uppercase\expandafter{\romannumeral2}}}(\bm{r})|_{x=L^{-}}= \Psi_{\mathbf{\uppercase\expandafter{\romannumeral3}}}(\bm{r})|_{x=L^{+}}.
\end{eqnarray}

The current density operator in the normal WSM region can be derived via
$\bm{J}=\frac{-i}{\hbar}[\bm{r},H_{BdG}]$.
The $x$-component is
\begin{eqnarray}
  J_{x}=\tau_{z}(v_{1}\sigma_{0}+v_{2}\sigma_{x}),
\end{eqnarray}
where $\tau_{z}$ is the $z$-component of Pauli matrix which acts in the electron-hole space.
Then the retro and specular AR coefficients $A_1$ and $A_2$ are evaluated by
\begin{eqnarray}
  A_{1/2}=\left|\frac{\langle\Psi_{h\pm}|J_{x}|\Psi_{h\pm}\rangle}{\langle\Psi_{e-}|J_{x}|\Psi_{e-}\rangle}\right||r_{1/2}|^{2}.
\end{eqnarray}
And the normal and specular ET coefficients $T_1$ and $T_2$ are evaluated by
\begin{eqnarray}
   T_{1/2}=\left|\frac{\langle\Psi_{e\mp}|J_{x}|\Psi_{e\mp}\rangle}{\langle\Psi_{e-}|J_{x}|\Psi_{e-}\rangle}\right||t_{1/2}|^{2}.
\end{eqnarray}
Because of the conservation of current, we have $ A_{1} + A_{2} + T_{1} + T_2=1$.


\begin{figure}
\begin{center}\scalebox{0.33}{\includegraphics{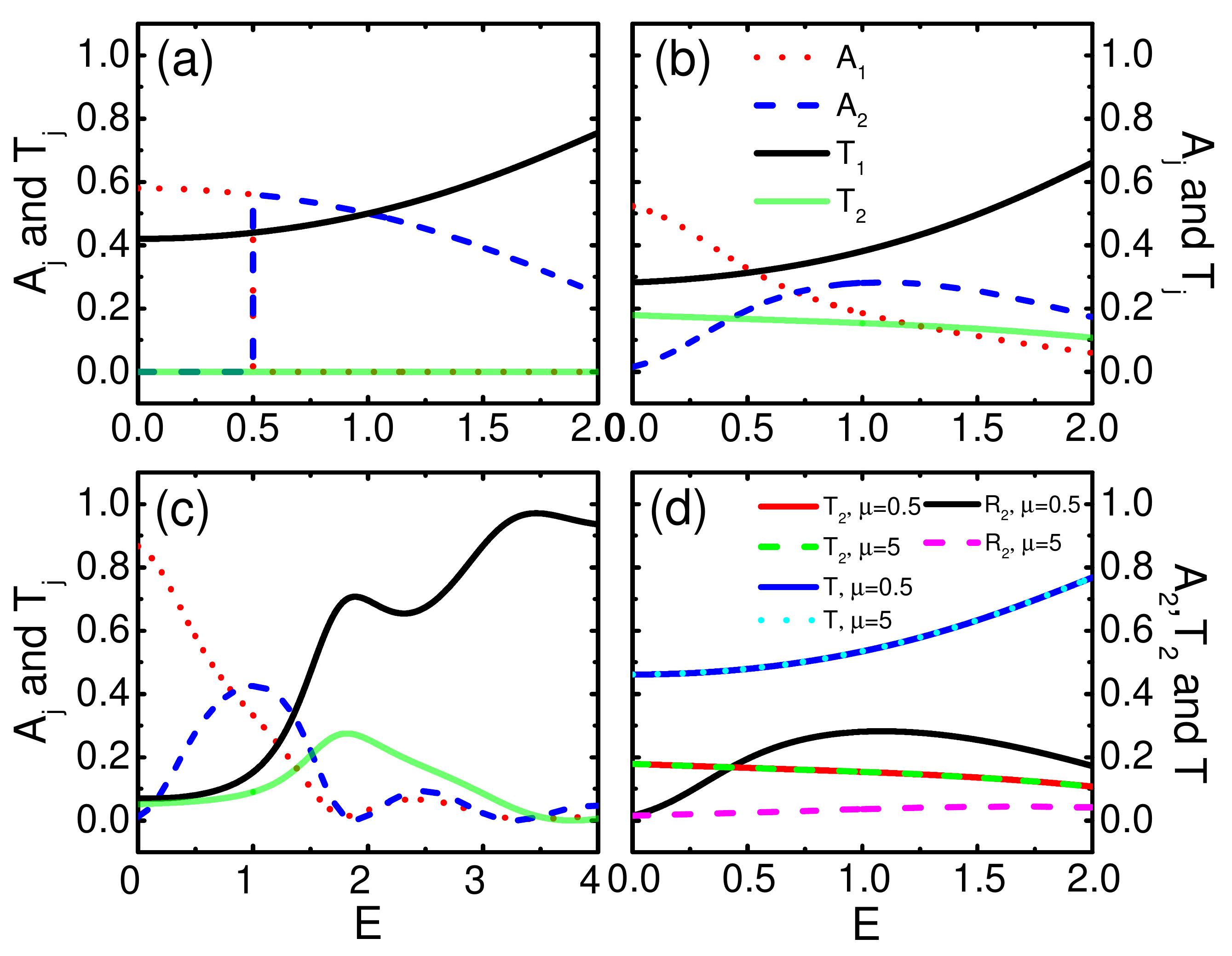}}
\caption{ (a-c) AR and ET coefficients as functions of incident energy $E$. (a) $\theta=0$, $L=\xi$, (b) $\theta=0.1\pi$ $L=\xi$, (c) $\theta=0.1\pi$ $L=2\xi$. (d) The specular AR coefficient $A_{2}$, specular ET coefficient $T_{2}$ and total transmission coefficient $T=T_1+T_2$ for chemical potential $\mu=0.5$ and $\mu=5$.  Parameters : (a)-(d) $k_z=0$, $v_1=2$, $v_2=1$, $U=100$, $\Delta = 1$; (a)-(c) $\mu=0.5$.
\label{fig3}}
\end{center}
\end{figure}

\subsection{Numerical results}

In Fig.~\ref{fig3}, we calculate the amplitudes of double ARs and double ETs, respectively. Fig.~\ref{fig3} (a) show the normal incidence case with junction length $L=\xi=\hbar(v_{1}-v_{2})/\Delta$. For the chemical potential, it is taken to be $\mu = 0.5$, which can be tuned by adjusting the gate voltage in experiment. One can find that the specular ET is forbidden, i.e., $T_2=0$, and the incident electron is Andreev reflected in the $A_1$ or $A_2$ mode for $E<\mu$ and $E>\mu$, respectively. However, the retro and specular modes coincide in the normal incident case to contribute identically to the quantum transport processes. Next, with the increase of incident energy, the ETs are enhanced while ARs are weakened monotonically in the energy regime $E<2\Delta$.
In the oblique incidence case shown in Fig.~\ref{fig3} (b)-(c), the four transport processes co-contribute to the transport through this junction. The retro AR coefficient $A_{1}$ decreases monotonically when the incident energy increases in the SC-gap regime $E<\Delta$, while the specular AR $A_2$ and normal ET coefficient $T_1$ exhibit increments in this regime, for junctions with length $L=\xi$ and $2\xi$ [See Fig.~\ref{fig3} (b)-(c)].
However, the specular ET coefficient $T_2$ can decrease or increase in the cases of $L=\xi$ [Fig.~\ref{fig3} (b)] and $L=2\xi$ [Fig.~\ref{fig3} (c)] in the low-energy regime $E<\Delta$.
In the regime $E>\Delta$, all these scattering processes exhibit an oscillatory behaviors due to the coherent tunnelling determined by the standing-wave condition. For the specular ET $T_2$, the valley (peak) of the oscillation emerges outside the gap regime leading to the increase (decrease) behavior in the gap regime for $L=\xi$ and $2\xi$, respectively. In
Fig.~\ref{fig3} (d), we present the total ET coefficient $T=T_1+T_2$, the specular ET coefficient $T_2$ and the specular AR coefficient $A_2$ as functions of the incident energy for chemical potentials $\mu=0.5$ and $5$. It can be clearly found that the change of chemical potential does not affect the ET coefficients $T$ and $T_1$, but only alters the relative weight of retro and specular ARs.

\begin{figure}
\begin{center}\scalebox{0.46}{\includegraphics{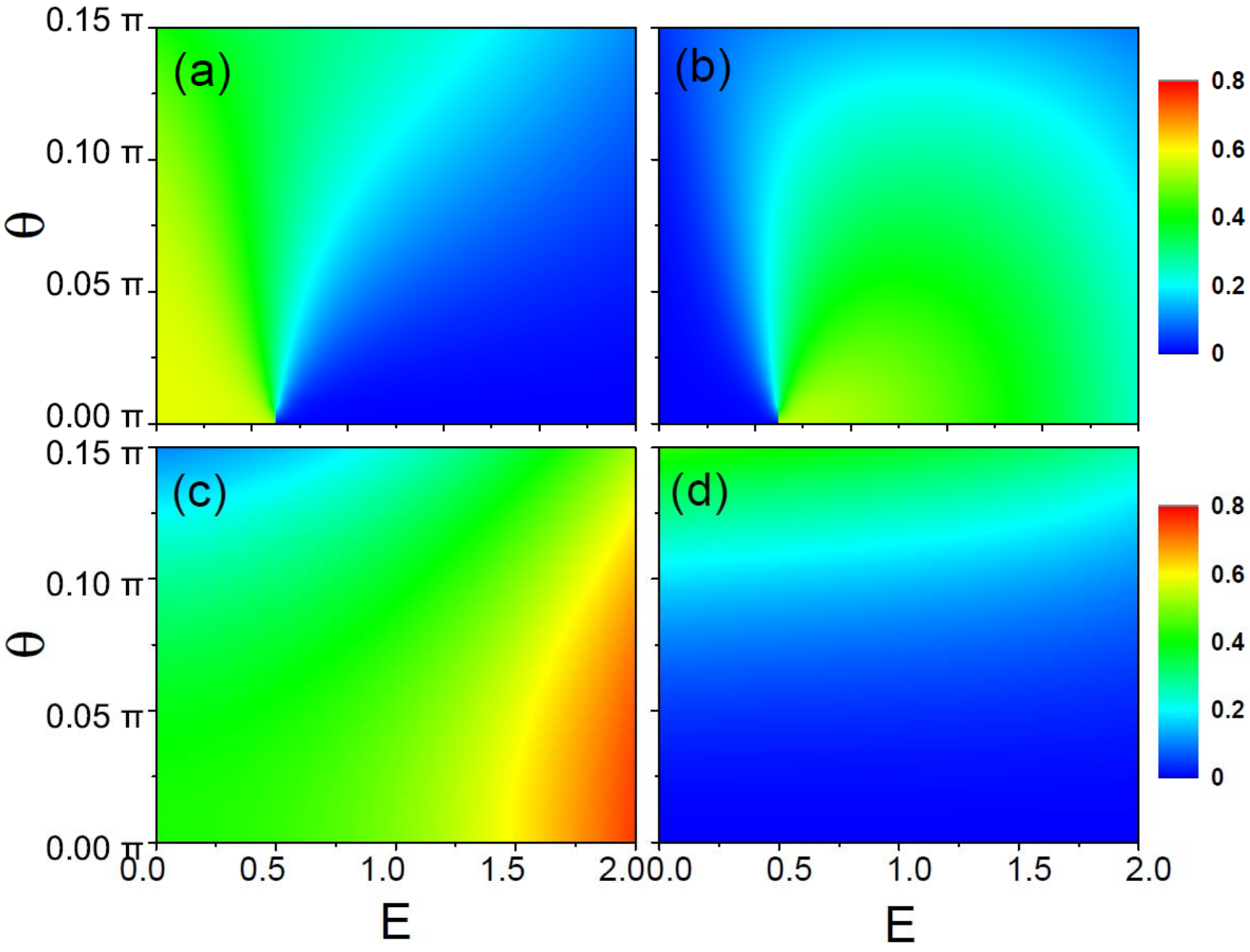}}
\caption{The incident angle and energy dependence of (a) retro AR $A_1$, (b) specular AR $A_2$, (c) intraband ET $T_1$, and interband ET $T_2$, respectively. Parameters: $v_{1}=2$, $v_{2}=1$, $L=\xi$, $U=100$, $\Delta=1$, $\mu=0.5$.
\label{fig4}}
\end{center}
\end{figure}


In what follows, we calculate the dependence of both double AR and double ET coefficients on incident angle $\theta$ and energy $E$, as shown in Fig.~\ref{fig4}. The incident angle is related to the group velocity by the formula $ {\rm tan} \theta = \frac{v_y}{v_x}$. By the velocity expressions in Eq.~(\ref{EqVg}), we know that $\theta$ has a upper limit $\theta_c = {\rm arctan}\frac{v_2}{\sqrt{v_1^2-v_2^2}}=0.17\pi$, since the equi-energy surface is a hyperbola in the WSM region. Next, in Fig.~\ref{fig4} (a) it shows that the retro AR $A_1$ decreases or increases monotonically with the increment of incident angle $\theta$ for electrons with incident energy $E<\mu$ and $E>\mu$, respectively.
However, compared with $A_1$, the specular AR $A_2$ exhibits an opposite dependence on $\theta$, as displayed in Fig.~\ref{fig4} (b). With respect to the double ETs, the results in Fig.~\ref{fig4} (c)-(d) show that $T_1$ and $T_2$ will decrease or increase monotonically if $\theta$ is increased in the whole energy regime $0<E<2\Delta$. Therefore, according to these results, we can find that the retro AR $A_1$ contributes mainly in the $E<\mu$ regime while specular AR $A_2$ mainly in the $E>\mu$ regime. This suggests that the AR can be tuned via a gate voltage. Alternatively, the intraband ET $T_1$ contributes almost for all incident energy and angles, while the specular ET $T_2$ occurs mainly for large incident angles.

\begin{figure}
\begin{center}\scalebox{0.4}{\includegraphics{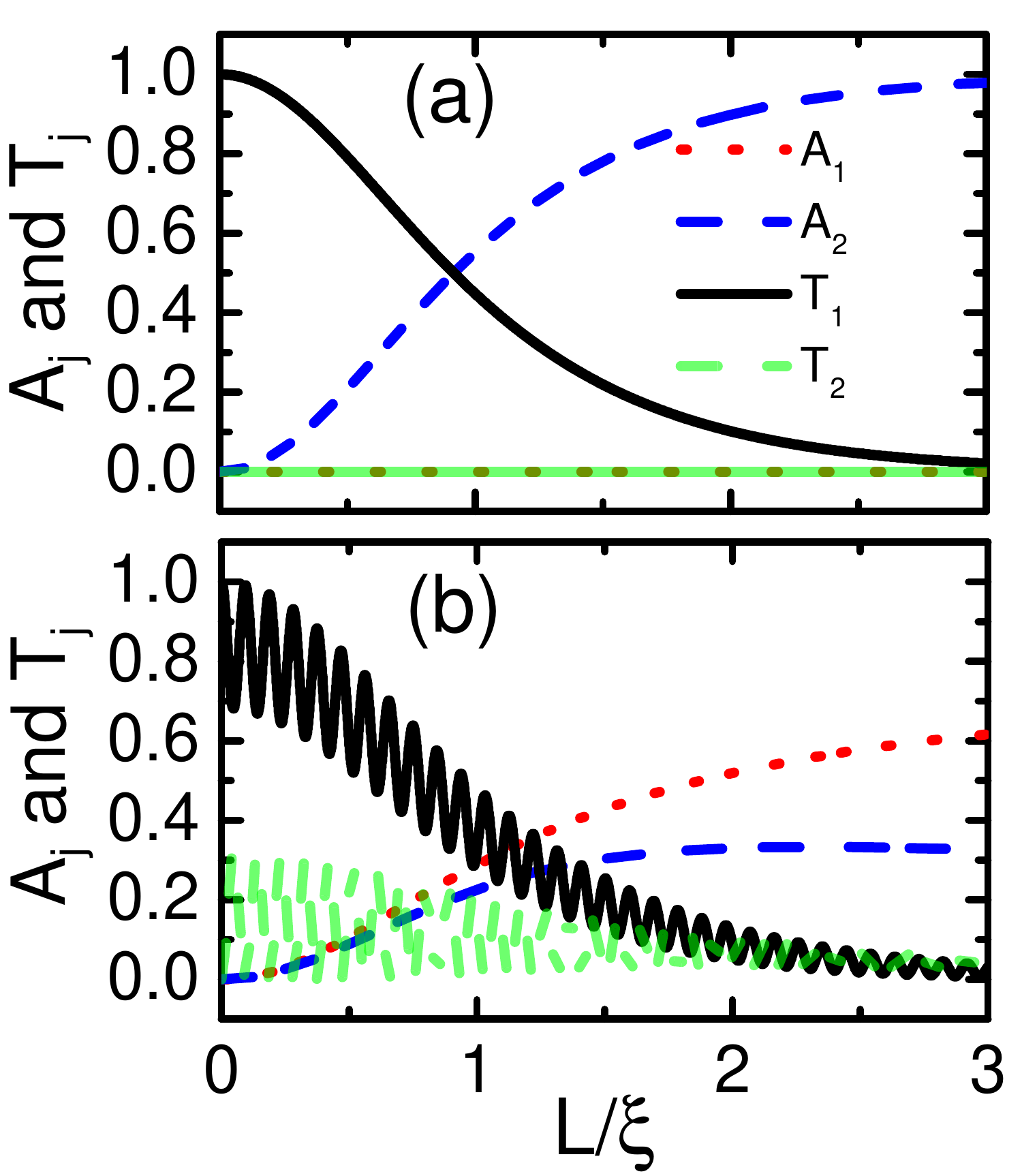}}
\caption{ AR and ET coefficients as functions of the SC region length $L$ with incident angle (a) $\theta=0$ and (b) $\theta=0.1\pi$. Parameters: $v_{1}=2$, $v_{2}=1$, $E=0.6$, $\mu=0.5$, $U=100$, $\Delta=1$.
\label{fig5}}
\end{center}
\end{figure}

In Fig.~\ref{fig5}, we study the dependence of AR and ET coefficients on the length of the SC region. Both the normal and oblique incidences [See Fig.~\ref{fig5} (b)] are considered, by taking the incident energy to be $E=0.6$.
Firstly, in Fig.~\ref{fig5} (a) one can find that in the normal incident case, only one AR mode and one ET mode appear, respectively, i.e., $A_2$ and $T_1$. This is exactly consistent with the result in Fig.~\ref{fig3} (a) where both $A_1$ and $T_2$ modes vanish for $E>\mu$.
In the short-junction limit $L\rightarrow0$, only the intraband ET $T_1$ is allowed with its amplitude $T_1=1$, just as expected, since no scattering potential exists. With the increase of $L$, ET is weakened gradually while AR enhanced. However, in the long-junction limit, ET is suppressed and the complete AR takes place, in accordance with the result of type-II WSM-SC junction~\cite{double}.
Next, in the oblique incidence case, both retro and specular AR coefficients, $A_{1}/A_{2}$, increase monotonically until reach the saturation value, with the increase of junction length. In contrast, the ET coefficients will decrease accompanied by oscillation. As a consequence, the peak of the normal-ET curve meets the specular-ET valley, in the short junction limit $L\rightarrow0$. The two ETs share an identical oscillation period, which can be approximated by ${\cal L} = \frac{2\pi}{|k_{x1}-k_{x2}|}$ determined from the standing wave condition. $k_{x1}$ ($k_{x2}$) is the real component of the wave vector of left (right) moving mode in the SC region. Most interestingly, there exists a phase shift of $\pi$ between normal and specular ETs, which vanishes the oscillation of the total transmission coefficient $T=T_1+T_2$.

\section{conductance}

\begin{figure}
\begin{center}\scalebox{0.4}{\includegraphics{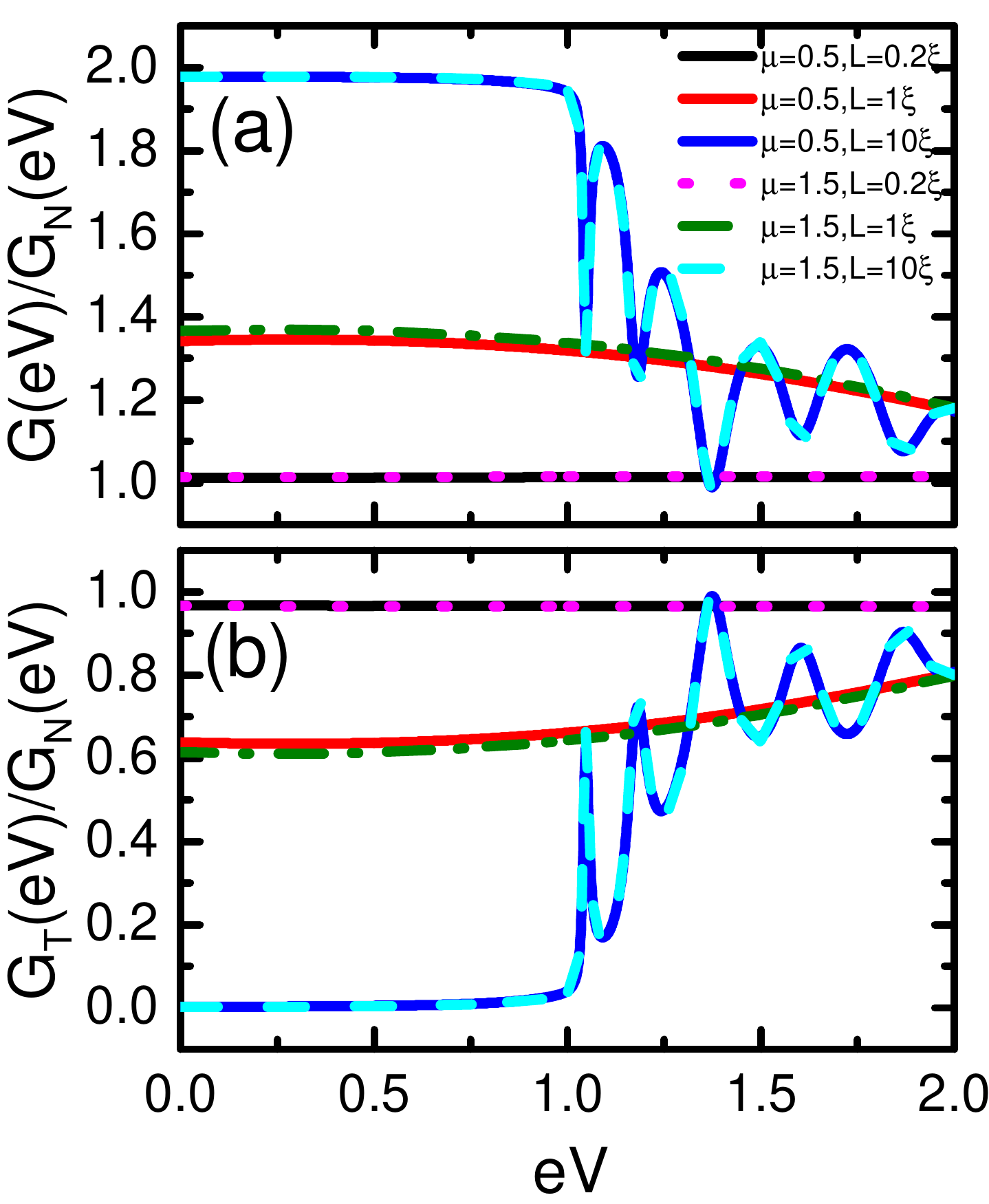}}
\caption{The conductance $G$ and its ET component $G_{T}$ as a function of the bias $eV$ for several chemical potentials and SC region size $L$ [(a) and (b)]. Parameters: $v_{1}=2$, $v_{2}=1$, $U=100$, $\Delta=1$, $q_{m}=10$.
\label{fig6}}
\end{center}
\end{figure}

Following the previous analysis of double ARs and double ETs, we investigate the properties of differential conductance in this section. We apply a bias voltage to the left WSM region, while the SC region and the right WSM region are grounded. Thus, the differential conductance can be calculated by the well-known BTK formula\cite{double,BTK}:
\begin{small}
\begin{eqnarray}%
G(eV)=\frac{e^{2}S}{\pi^{2}h} \sum_{m} \int\int dk_y dk_z [1 +A^{(m)}(k_y,k_z,eV)  ],
\end{eqnarray}\end{small}
where $S$ is the cross-sectional area of the junction, $m=\pm$ is to distinguish the contribution of transport processes due to incident modes $\psi_{e\pm}$, and $A^{(m)}=A_1^{m}+A_2^{m}$ is the total AR coefficient. The integration over the wave vectors is constrained near the Weyl node with a cut-off $q_m$, i.e., $\sqrt{(k_y^2+k_z^2)}<q_m$. The contributions of ETs and ARs to the conductance, i.e., $G_T$ and $G_A$, can be written as
\begin{small}
\begin{eqnarray}%
 &&G_T(eV)=\frac{2e^{2}S}{\pi^{2}h}\int\int dk_y dk_z  T(k_y,k_z,eV),\nonumber\\
 &&G_A(eV)=\frac{2e^{2}S}{\pi^{2}h}\int\int dk_y dk_z 2A(k_y,k_z,eV).
\end{eqnarray}\end{small}
If we denote $G_{N}(eV)=\frac{2e^{2}S}{\pi^{2}h}\cdot\pi q_{m}^{2}$, the conductance of a normal junction, there will be
\begin{eqnarray}\label{EqGAT}
 & G+G_T=2G_N,\nonumber\\
 & G_A = 2G - 2G_N.
\end{eqnarray}
 This indicates that it is impossible to enhance both the total conductance and ET conductance at the same time.

The conductance $G$ and its ET component $G_T$ as functions of bias for several chemical potential $\mu$ and junction length $L$ are shown in Fig.~\ref{fig6} (a) and (b), respectively.
It can be obviously seen that $G$ and $G_T$ decrease or increase with the enhancement of the  bias voltage, respectively. This is consistent with the result in Eq.~\ref{EqGAT} and the previous analyses of ET coefficients, which can be increased by the enlargement of incident energy. Besides, it shows that the conductance is independent of chemical potential $\mu$, such a result can shown by understood by observing the properties of scattering coefficients $A$ and $T$.
On the other hand, the conductance can be affected by the junction length to a great extent. Namely, the longer junction will induce the weaker ET conductance while the larger AR conductance and therefore the larger total conductance, since the ET (AR) coefficients decreases (increases) in this process [See Fig.~\ref{fig5}].
In the long-junction limit, $G_T$ will vanish, leading to the occurrence of the perfect AR in the SC-gap regime $E<\mu$. One can then find the result that $G/G_N=2$. If the bias is further increased, $G$ ($G_T$) decreases (increases) gradually with an oscillation, due to the coherent tunnelling determined by the standing-wave condition. However in the short-junction limit, both $G$ and $G_T$ are approximately equal to $G_N$ in the SC-gap regime, because of the vanishing of AR.
In addition, we notice that the conductance here has a large magnitude, the reason is the consideration of the large momentum is mismatched between the normal and SC region, which are identical to an effective scattering potential. In fact, this is exactly the forbidden of NR in our model leads to this phenomenon.

\section{Summary}
In Summary, we have studied the transport property of the NSN junction based on type-II WSM, and predicted the novel phenomenon of double ARs and double ETs.
It means that four scattering processes for incident electrons coexist in this system, i.e., the retro and specular ARs, the normal and specular ETs. However, the NR and crossed AR are forbiden. The retro and specular AR modes are symmetric about the normal of the interface but with different amplitudes, which is also the case for normal and specular ET processes.
The dependences of the amplitudes on the incident angle and energy have been studied in detail with the effect of chemical potential and junction length considered.
In addition, we have studied the differential conductance. It is found that the conductance is independent of chemical potential of the WSM region, suggesting its robustness. The normalized conductance will increase as enlarging junction length. The conductance has been formed to have a large value considering the scattering potential due to the momentum mismatch between the normal and SC region.

\section*{Acknowledgments}
This work is supported by the Liaoning BaiQianWan Talents program, the National Natural Science Foundation of China (Grant No. 11747122), the Natural Science Foundation of Shandong Province (Grant No. ZR2018PA007), and the Doctoral Foundation of University of Jinan (Grant No. 160100147).

\begin{appendix}
\section{eigenvectors}
In this appendix, we give the eigenvectors in the WSM and SC regions.

The BdG equation in WSM region is represented as
\begin{eqnarray}
  \left[
\begin{array}{cc}
H_{+}(\bm{k})-\mu & 0 \\
0  & -H_{+}(\bm{k})+\mu
\end{array}
\right]\Psi
=E\Psi
\end{eqnarray}
Two incident and two reflected modes with energy $E$ and wave vector $k_y$ and $k_z$ can be given as\cite{double}
\begin{small}
\begin{eqnarray}
&&\Psi_{e+}({\bf r})=
\left[
\begin{array}{cc}
k_{+}+k_{z}\\
k_{x+}+ik_{y}\\
0\\
0
\end{array}
\right]
\exp(ik_{x+}x+ik_{y}y+ik_{z}z),
\nonumber\\
&&\Psi_{e-}({\bf r})=
\left[
\begin{array}{cc}
-k_{-}+k_{z}\\
k_{x-}+ik_{y}\\
0\\
0
\end{array}
\right]
\exp(ik_{x-}x+ik_{y}y+ik_{z}z),
\nonumber\\
&&\Psi_{h+}({\bf r})=
\left[
\begin{array}{cc}
0\\
0\\
-k_{+}'+k_{z}\\
k_{x+}'+ik_{y}
\end{array}
\right]
\exp(ik_{x+}'x+ik_{y}y+ik_{z}z),
\nonumber\\
&&\Psi_{h-}({\bf r})=
\left[
\begin{array}{cc}
0\\
0\\
k_{-}'+k_{z}\\
k_{x-}'+ik_{y}
\end{array}
\right]
\exp(ik_{x-}'x+ik_{y}y+ik_{z}z). \nonumber \\\label{e1}
\end{eqnarray}
\end{small}
In Eq.(\ref{e1}), the wave vectors and their $x$- and $y$-components of each energy band are respectively given as
\begin{small}
\begin{equation}
\begin{split}
  &k_{x\pm}=\frac{v_{1}(E+\mu)\mp v_{2}\sqrt{(E+\mu)^{2}+\hbar^{2}(v_{1}^{2}-v_{2}^{2})
  (k_{y}^{2}+k_{z}^{2})}}{\hbar(v_{1}^{2}-v_{2}^{2})},\nonumber \\
  &k_{x\pm}'=\frac{-v_{1}(E-\mu)\pm v_{2}\sqrt{(E-\mu)^{2}+\hbar^{2}(v_{1}^{2}-v_{2}^{2})
  (k_{y}^{2}+k_{z}^{2})}}{\hbar(v_{1}^{2}-v_{2}^{2})},\nonumber\\
  &k_{\pm}=\sqrt{k_{x\pm}^{2}+k_{y}^{2}+k_{z}^{2}}, k_{\pm}'=\sqrt{k_{x\pm}'^{2}+k_{y}^{2}+k_{z}^{2}},
\end{split}
\end{equation}
\end{small}
where $k_{x\pm}$ and $(k_{x\pm}')$ are corresponding $x$-component of wave vectors, shown in Fig.~\ref{fig_s2}.

The BdG equation of the SC region is given in Eq.~\ref{EqHBdG}. In the large-$U$ limit\cite{AR-G1}, the four excited modes with energy $E>0$ and wave vector $k_y$ and $k_z$ can be approximated as,\cite{double}
\begin{small}
\begin{eqnarray}
&&\Psi_{a}({\bf r})\!=\!
\left[
\begin{array}{cc}
e^{-i\beta}\\
e^{-i\beta}\\
1\\
1
\end{array}
\right]
\exp(ik_{L-} x+ik_{y}y+ik_{z}z),\nonumber\\
&&\Psi_{b} ({\bf r})\!=\!
\left[
\begin{array}{cc}
e^{i\beta} \\
e^{i\beta} \\
1\\
1
\end{array}
\right]
\exp(ik_{L+} x+ik_{y}y+ik_{z}z),
\nonumber\\
&&\Psi_{c} ({\bf r})\!=\!
\left[
\begin{array}{cc}
e^{-i\beta}\\
-e^{-i\beta}\\
1\\
-1
\end{array}
\right]
\exp(ik_{R-} x+ik_{y}y+ik_{z}z),
\nonumber\\
&&\Psi_{d} ({\bf r})\!=\!
\left[
\begin{array}{cc}
e^{i\beta}\\
-e^{i\beta}\\
1\\
-1
\end{array}
\right]
\exp(ik_{R+} x+ik_{y}y+ik_{z}z).
\end{eqnarray}
\end{small}
$\Psi_{b/d}$ ($\Psi_{a/c}$) are right (left) moving modes with positive slope as shown in Fig.~\ref{fig_s2}. And the parameters are
\begin{eqnarray}\label{EqPS}
&&\beta=\left\{
\begin{array}{lll}
\arccos(E/\Delta)       &      & {\textrm{if}\quad E      <      \Delta},\\
-i\,\textrm{arcosh}(E/\Delta)       &      & {\textrm{if}\quad E     >      \Delta},
\end{array} \right.
\nonumber\\
&&k_{xL-} = k_{x1}-i\tau_{1},\
k_{xL+} =k_{x1}+i\tau_{1},
\nonumber\\
&&k_{xR-} =k_{x2}-i\tau_{2},\
k_{xR+} =k_{x2}+i\tau_{2},
\nonumber\\
&&k_{x1}\simeq\frac{U}{v_{1}+v_{2}},\
k_{x2}\simeq\frac{U}{v_{1}-v_{2}},
\nonumber\\
&&\tau_{1}=\frac{\Delta\sin{\beta}}{\hbar(v_{1}+v_{2})},\
\tau_{2}=\frac{\Delta\sin{\beta}}{\hbar(v_{1}-v_{2})}.
\end{eqnarray}
where $k_{xL\pm}$ and $(k_{xR\pm}')$ are corresponding to the $x$-component of wave vectors, shown in Fig.~\ref{fig_s2}.

\end{appendix}

\clearpage

\bigskip

\end{document}